\documentstyle[12pt]{article}

\newcommand \ra {\rightarrow}

\newcommand{\be}{\begin{equation}}
\newcommand{\ee}{\end{equation}}
\newcommand{\bea}{\begin{eqnarray}}
\newcommand{\eea}{\end{eqnarray}}
\newcommand \boundary {\partial}
\newcommand{\zed}{{\mathbf{Z}}}

\newcommand{\dif}{\Delta}

\newcommand \s {\sigma}

\def\reff#1{(\ref{#1})}
\newtheorem{theorem}{Theorem}

\begin{document}

\title{Expansions for Droplet States in the Ferromagnetic XXZ Heisenberg 
Chain} 

\author{
\\Tom Kennedy
\\Department of Mathematics
\\University of Arizona
\\Tucson, AZ 85721
\\ email: tgk@math.arizona.edu
\bigskip
}

\maketitle

\begin{abstract}
We consider the highly anisotropic ferromagnetic spin 1/2 Heisenberg 
chain with periodic boundary conditions. 
In each sector of constant total z component of the spin, we
develop convergent expansions for the lowest band of 
eigenvalues and eigenfunctions. These eigenstates describe 
droplet states in which the spins essentially form a single linear
droplet which can move. Our results also give a convergent expansion 
for the dispersion relation, i.e., the energy of the droplet as a function
of its momentum. The methods used are from \cite{DKa} and \cite{DKb}, 
and this short paper should serve as a pedagogic introduction to 
those papers. 

\end{abstract}

\newpage

\section{Introduction}

We consider the spin $1/2$ ferromagnetic XXZ Heisenberg chain. 
With periodic boundary conditions its Hamiltonian is 
\be
H = -\sum_{j=1}^N [\s^z_j \s^z_{j+1} + \epsilon \s^x_j \s^x_{j+1} 
+ \epsilon \s^y_j \s^y_{j+1}] 
\ee
We will always assume that $\epsilon$ is small, so we are in the Ising-like
regime. This Hamiltonian commutes with the total $z$ component of the 
spin, so one can study it in a sector with a fixed total $z$ component of 
spin. The lowest energy states in each such
sector are droplet states in which the down spins essentially
form a single droplet. This droplet can move around, so the lowest
part of the spectrum in each sector should be a band of continuous spectrum.
Much of this picture was proved by Nachtergaele and Starr \cite{NS}.
In particular, they showed that the eigenvalues of these droplet states
lie in an interval whose width is of order $\epsilon^m$ where $m$ is the 
number of spins in the droplet, and there is 
a gap between this band and the rest of the spectrum. They gave an 
explicit description of the subspace spanned by these droplet eigenstates
which becomes exact in the limit that the size of the droplet goes 
to infinity. 

We will show that methods from \cite{DKa} and \cite{DKb} 
can be used to construct 
convergent expansions for the eigenstates in this lowest band in 
each sector with a fixed number of down spins. 
Because the translation-invariant ground states of the 
ferromagnetic XXZ chain are trivial,
the estimates in this paper are considerably simpler than those in
\cite{DKa} and \cite{DKb}. Thus this short paper should serve as a 
pedagogic introduction to those papers. 
At first glance it may be surprising that one can construct convergent 
expansions for eigenstates that will become part of continuous spectrum 
in the limit that the length of the chain goes to infinity. 
We avoid this problem by using the 
fact that with periodic boundary conditions the eigenfunctions have a 
definite momentum. Within the subspace of a given momentum the 
eigenvalue we want to study is isolated, and a convergent expansion
is possible. The results of \cite{NS} are valid for $|\epsilon|<1$,
while our results are valid only for small $\epsilon$. So our results
have a smaller range of applicability, but give more detail about 
the droplet eigenstates and eigenvalues.

Before we consider the droplet states, we will first consider the chain
with the periodic boundary conditions replaced by boundary
conditions that favor the spin to be up at one end of the chain and down
at the other end of the chain. Such boundary conditions force 
a kink or interface into the chain. 
For a generic Hamiltonian one would not expect 
this interface to be stable. Roughly speaking it would 
be equally likely to be anywhere along the chain. However, for the 
model considered here the conservation of total $z$ component of spin
essentially fixes the location of the kink. 
The stability of this interface 
was proved independently by Alcaraz, Salinas and Wreszinski \cite{ASW} and
Gottstein and Werner \cite{GW}. Alternative proofs of the 
stability of this interface were given in \cite{BCN}, 
by using the path integral representation of interface states, 
and in \cite{BM}. 
Other work on these interface states includes 
\cite{KN1}, \cite{KN2}, \cite{KNS1}, \cite{KN3}, \cite{Mat1}, and \cite{Mat2}.
Our discussion of these interface states
does not contain any new results and is 
intended as an introduction to the methods used for developing the 
expansion of the droplet states.

We will always work on finite chains with estimates independent of the 
length of the chain. Of course, one obtains continuous spectra only in the 
infinite length limit. It would be interesting to show that in the 
infinite length limit, defined for example via the GNS construction,
there is indeed a band of continuous spectrum corresponding to the motion of 
the droplets. 

This paper only concerns the one-dimensional XXZ ferromagnet. There are 
many interesting results on interfaces in the higher dimensional model
(see \cite{nac} for a review), 
but we are not aware of any results on droplet states in the higher
dimensional model. In two dimensions, if $m$, the number of spins in the 
droplet, is a perfect square, then the ground states of the $\epsilon=0$ 
Hamiltonian are just droplet states in which the droplet is a square. 
So the $\epsilon=0$ ground state is unique up to translations.
In these sectors it may be possible to use the methods of this paper
to contruct the droplet eigenstates. (A similar remark applies in higher
dimensions.) For other sectors the $\epsilon=0$ 
ground states are typically more degenerate, and the methods of 
this paper would need to be combined with some form of degenerate 
perturbation theory.

\section{Kink states}

In this section we construct convergent expansions for the ``kink states''
which contain a single domain wall. These states have been studied
extensively, and there are no new results in this section. This section 
serves as an introduction to our expansions for the droplet states 
in the next section. The expansion for the kink states is similar to that
for the droplet states, but simpler since one does not have to deal 
with the freedom of the droplets to move about. 

We consider the ferromagnetic XXZ Heisenberg chain on $N$ sites
with the following boundary conditions.
\be
H = \sum_{j=1}^{N-1} [1-\s^z_j \s^z_{j+1} - \epsilon \s^x_j \s^x_{j+1}
- \epsilon \s^y_j \s^y_{j+1}]
+ A (1+\s_1^z)  + B (1-\s_N^z) 
\label{ham_open}
\ee
The constants $A$ and $B$ will be positive, so the boundary conditions
favor the spin down state at site $1$ and the spin up state at site $N$. 
If we consider the sector with $m$ down spins, then the above Hamiltonian
with $\epsilon=0$ has a unique ground state given by putting 
the $m$ down spins at sites $1$ to $m$ and the up spins at the other sites.
For a set $X$ of sites we define $|X\rangle$ to be the state with 
down spins at the sites in $X$ and up spins everywhere else. 
So the $\epsilon=0$ ground state is $|\{1,2,\cdots,m\}\rangle$.
The set $\{1,2,\cdots,m\}$ will appear throughout 
our equations, so we denote it by 
\be
M=\{1,2,\cdots,m\} .
\ee
When $\epsilon$ is not zero, the ground state in the sector with $m$ 
down spins will be a perturbation of $|M\rangle$. The coefficient of 
$|Y\rangle$ should get smaller as $Y$ gets ``farther'' from $M$. 
Thus it is natural to write $Y$ as a modification of $M$. For sets of 
sites $X$ and $Y$ we denote by $X \dif Y$ the set of sites which 
belong to exactly one of $X$ and $Y$. So
\be
X \dif Y = (X \cup Y) \setminus (X \cap Y)
\ee
Any state in the sector with $m$ down spins can be written as 
\be
\Psi = \sum_{X:m} e(X) \, |X \dif M\rangle
\ee
for some coefficients $e(X)$. The constraint $X:m$ means that we only 
sum over sets $X$ such that $|X \dif M \rangle$ has $m$ down spins. 
This means that $X \dif M $ has $m$ elements, or equivalently that 
$X \cap M$ and $X \cap M^c$ have the same number of elements. 
For example, $X$ could be $\{m,m+1\}$ in which case $|M \dif X\rangle$ 
is the state that occurs at first order in perturbation theory. 

When $\epsilon=0$, $|X \dif M\rangle$ is an eigenstate for all $X$. 
We let $\lambda(X)$
denote its eigenvalue. It equals $2$ times the number of 
nearest neighbor pairs of sites for which the spins in 
$|X \dif M\rangle$ are anti-parallel plus boundary terms. 
The boundary terms are $+2A$ if $1 \in X$ and 
$+2B$ if $N \in X$. For the groundstate, $X=\emptyset$, we have 
$\lambda(X)=2$.
If we have $A \ge 1$ and $B \ge 1$, then for 
all other $X$ with $X:m$ we have $\lambda(X) \ge 6$. 
(Note that there are other $X$ with $\lambda(X)=2$, e.g., $X=\{m+1\}$,
but they belong to sectors with a different number of down spins.)
If $A$ and $B$ are positive, but possibly less than $1$, then the 
lower bound of $6$ must be replaced by the minimum of $6$ and 
$2+2A+2B$. The following proof applies in this case with slight 
modifications. To keep things as simple as possible we will assume 
$A$ and $B$ are both at least $1$. (In the more general case, just how
small $\epsilon$ would need to be would depend on $A$ and $B$.)

We now consider the Hamiltonian with nonzero $\epsilon$.
The term
$\s^x_j \s^x_{j+1} + \s^y_j \s^y_{j+1}$
can be written as  $\s^x_j \s^x_{j+1}(1 - \s^z_j \s^z_{j+1}) $.
So if the spins at $j$ and $j+1$ are parallel, it gives zero and if they
are anti-parallel it flips these two spins and multiplies the result by 2. 
We let $\boundary(X \dif M)$ denote the set of bonds $<j,j+1>$ 
such that the spins at $j$ and $j+1$ in $|X \dif M\rangle$ are different.  
One can think of $\boundary(X \dif M)$ as the set of domain walls in 
the spin configuration.
Then
\be 
H \Psi = \sum_{X:m} \lambda(X) e(X) \, |X \dif M\rangle
+ 2 \epsilon \sum_{X:m} e(X) \sum_{<j,j+1> \in \boundary(X \dif M)} 
|\{j,j+1\} \dif X \dif M\rangle
\ee
Note that $<j,j+1> \in \boundary(X \dif M)$ if and only if  
$<j,j+1> \in \boundary(\{j,j+1\} \dif X \dif M)$. So we can do a change of 
variables $X \ra X \dif \{j,j+1\}$ in the last term and rewrite this as 
\be 
H \Psi = \sum_{X:m} \lambda(X) e(X) \, |X \dif M\rangle
+ 2 \epsilon \sum_{X:m} \quad
\sum_{<j,j+1> \in \boundary(X \dif M)} e(X \dif \{j,j+1\}) \, |X \dif M\rangle
\ee
When $\epsilon=0$, the ground state energy is $2$, so we will write 
the ground state energy for nonzero $\epsilon$ in the form $2+E$, i.e,
we look for a solution of $H \Psi = (2+E) \Psi$. 
Equating the coefficients of 
$|X \dif M \rangle$, we have for $X$ with $X:m$, 
\be
\lambda(X) e(X) 
+ 2 \epsilon \sum_{<j,j+1> \in \boundary(X \dif M)} e(X \dif \{j,j+1\})
= (2+E) e(X)
\ee

We take $e(\emptyset)=1$. (At first we cannot be sure this is possible since
$e(\emptyset)$ could be zero for the groundstate. 
But if we can succeed in constructing an eigenstate with this condition and 
show it is the ground state, then this will justify the assumption.)
The unknowns in the above equation are $e(X)$ for non-empty $X$ with $X:m$
and $E$. For $X=\emptyset$ the above equation gives
\be
 E = 2 \epsilon \, e(\{m,m+1\})
\label{kink_fpea}
\ee
and for non-empty $X$ with $X:m$ the above equation can be written as
\be
e(X) = {1 \over \lambda(X)-2} \left[ -2 \epsilon 
\sum_{<j,j+1> \in \boundary(X \dif M)} e(X \dif \{j,j+1\}) + E e(X) \right]
\label{kink_fpeb}
\ee

We think of equations \reff{kink_fpea} and \reff{kink_fpeb} as a fixed point 
equation for the unknowns $E$ and $e(X)$, where $X$ ranges over non-empty
sets with $X:m$. 
We use $e$ to denote this collection of unknowns and write equations 
\reff{kink_fpea} and \reff{kink_fpeb} together as $F(e)=e$. 
We define a norm on the space of $e$'s by
\be 
||e||=|E| + \sum_{X:m, X \ne \emptyset} (\lambda(X)-2) |e(X)|
\ee
We will show that $F$ is a contraction in a small ball about the origin and
maps this ball back into itself. The existence of a solution then follows
by the contraction mapping theorem. 

At first glance it seems that $F(0)=0$, but it is not. This is because
$e(\emptyset)$ can appear in the right side of \reff{kink_fpeb}.
This happens when $X=\{m,m+1\}$ and $j=m$. This produces a ``constant'' 
term in $F(e)$, i.e., a term which does not depend on $e$. 
Denoting the terms in $F(0)$
by $e_0(X)$ we see that $e_0(\{m,m+1\})= -\epsilon/2$, and the other 
$e_0(X)$ and $E_0$ are all zero. We note, for later use, that 
$||F(0)||=2 \epsilon$.

To show that $F$ is a contraction we have 
\bea 
||F(e)-F(e')|| &\le& 2 \epsilon |e(\{m,m+1\})-e'(\{m,m+1\})|
\nonumber \\
&+& 2 \epsilon 
\sum_{X:m, X \ne \emptyset} \quad 
\sum_{<j,j+1> \in \boundary(X \dif M) }
|e(X \dif \{j,j+1\} ) - e'(X \dif \{j,j+1\})|
\nonumber \\
&+& \sum_{X:m, X \ne \emptyset} |E e(X) - E' e'(X)| 
\label{cbound}
\eea
We split this bound into two parts. The first term in the right side of 
the above is just what the second term would give when $X=\emptyset$. 
So the first two terms of the three terms in the right side of the above are
\bea 
&=& 
 2 \epsilon \sum_{X:m} \quad \sum_{<j,j+1> \in \boundary(X \dif M) } 
|e(X \dif \{j,j+1\} ) - e'(X \dif \{j,j+1\})|
\nonumber \\
&=& 
 2 \epsilon \sum_{X:m} \lambda(X) |e(X) - e'(X)|
\eea
where the last equality follows by a change of variables, 
$X \ra X \dif \{j,j+1\}$. For $X=\emptyset$, $|e(X) - e'(X)|=0$.
For all other $X$ with $X:m$ we have $\lambda(X) \ge 6$ and so 
$\lambda(X) \le {3 \over 2} (\lambda(X)-2)$. Thus the above is 
\be
\le 3 \epsilon ||e - e'||
\ee

The second part of \reff{cbound} is 
\bea
\sum_{X:m, X \ne \emptyset} | E e(X)-E'e'(X)|
&\le& 
\sum_{X:m, X \ne \emptyset} [ \, |E| \, \, | e(X)-e'(X)|
 + |E-E'| \, \, |e'(X)| \, ]
\nonumber \\
&\le & {1 \over 4} \sum_{X:m, X \ne \emptyset} (\lambda(X)-2) \, 
[|E| \, \, | e(X)-e'(X)| + |E-E'| \, \, |e'(X)| \, ]
\nonumber \\
&\le& {1 \over 4} ||e-e'|| \quad \max \{||e||,||e'||\}
\eea
where we have used $\lambda(X)-2 \ge 4$. 

Putting together these two bounds, we have shown for $e$,$e'$ with 
norm less than $\delta$ that 
\be 
||F(e)-F(e')|| \le (3 \epsilon + {1 \over 4} \max \{||e||,||e'||\}) ||e-e'||
\le (3 \epsilon + {\delta \over 4} ) ||e-e'||
\ee
Thus $F$ is a contraction on the ball of radius $\delta$ about the origin
if $\delta$ and $\epsilon$ are small enough. 
To see that it maps the ball of radius $\delta$ about the origin back into 
itself, we use 
\be 
||F(e)|| \le ||F(0)|| + ||F(e)-F(0)|| \le 2 \epsilon + 
(3 \epsilon+ {\delta \over 4})
 ||e-0|| < \delta 
\ee
if $||e||<\delta$ and $\epsilon$ is sufficiently small. 

By using a stronger norm we can derive decay properties for the coefficients
$e(X)$. Let $K$ be a large positive constant. Define 
\be 
||e||= |E| (K |\epsilon|)^{-2} + 
\sum_{X:m, X \ne \emptyset} (\lambda(X)-2) |e(X)| (K |\epsilon|)^{-w(X)}
\label{weighted}
\ee
We will assume that $\epsilon$ is small enough that $K |\epsilon| < 1$. 
$w(X)$ is a positive integer which is the lowest order in perturbation 
theory at which $e(X)$ gets a nonzero contribution. More precisely, we 
consider all sequences of sets $X_0,X_1, \cdots,X_n$ such that $X_0=\emptyset$,
$X_n=X$ and for $i=1,2,\cdots,n$, $X_i=X_{i-1} \dif \{j,j+1\}$ for 
some $j \in \boundary (X_{i-1} \dif M)$. Then $w(X)$ is the smallest $n$ for
which such a sequence exists. 

The previous estimates that proved the existence of a fixed point in our
original norm can be repeated with this new norm. One finds that the estimates
continue to hold provided $w(X)$ satisfies the following three 
properties.
\bea
&& w(\{m,m+1\})=1 
\nonumber \\
&& w(X \dif \{j,j+1\})\le w(X)+1
\nonumber \\
&& w(X) -2 \le w(X)
\nonumber \\
\eea
The first two properties follow easily from the definition of $w(X)$, while
the third is trivial. Note that the existence of a solution to the fixed 
point equation in this stronger norm implies that the coefficients $e(X)$
decay at least as fast as $(K |\epsilon|)^{w(X)}$. 

The expansion that we have developed can now be used to study a variety
of properties of the kink states. For example, the localization of the 
kink near the site $m$ follows from these estimates. One can consider the 
dependence of the ground state energy on the sector (the choice of $m$).
Most studies of the kink states used particular values of $A$ and $B$ for
which this energy is independent of the sector. Bach and Macris \cite{BM}
considered more general boundary conditions and showed that the 
difference between the ground state energies in different sectors were 
exponentially small in the length of the chain provided the kink is 
not near the boundaries. With more work it is probably possible to rederive
this result with our expansion. 
Our interest here is primarily in the droplet states, so we do not pursue 
this approach to the kink states any further. 

\section{Droplet states}

Now we consider the ferromagnetic XXZ Heisenberg chain with periodic 
boundary conditions.
For $N$ sites its Hamiltonian may be taken to be
\be
H = \sum_{j=1}^N [1-\s^z_j \s^z_{j+1}
- \epsilon \s^x_j \s^x_{j+1} - \epsilon \s^y_j \s^y_{j+1}] 
\label{ham}
\ee
Indices will always be taken to be periodic, e.g., $\s^z_{N+1}$
means $\s^z_1$.

We consider the sector with $m$ down spins. 
We continue to use the abbreviation 
\be
M=\{1,2,\cdots,m\}
\ee
For $\epsilon=0$ the ground states in this sector are $|M\rangle$ and 
its translates. When $\epsilon \ne 0$, the interface between sites
$m$ and $m+1$ and between sites $1$ and $N$ will spread out somewhat. 
States of the form 
$| M \dif X \rangle$ where $X$ is small and localized 
near $1$ and $m$ will make up the dominant part of the eigenstate. 
We continue to denote the constraint that $|X \dif M\rangle$ is in the 
sector with $m$ down spins by $X:m$. 
We now look for eigenstates with momentum $k$ in the form
\be
\Psi_k = 
\sum_{l=1}^N e^{ikl} \sum_{X:m} e(X) \, \, |(X \dif M)+l\rangle
\label{eigenfunction}
\ee
For a set of sites $Y$, we use $Y+l$ to denote the translate of the set 
by $l$ sites to the right. So $Y + l = \{i+l:i \in Y\}$.
This state has momentum $k$ in the sense that if $T$ is the operator of 
translation by one lattice site to the right, then 
$T \Psi_k = e^{-ik} \Psi_k$. 

As before, $\boundary X$ is 
the set of bonds such that one endpoint is in $X$ and the other 
endpoint is not in $X$. 
We define $n(X)$ to be the number of bonds in $\boundary (X \dif M)$. 
Then $2 n(X)$ is the eigenvalue of $|X \dif M \rangle$ when $\epsilon=0$. 
We have
\bea 
H \Psi_k &=& 
2 \sum_{l=1}^N e^{ikl} \sum_{X:m} e(X) 
 n(X) \, \, |(X \dif M)+l\rangle
\nonumber \\
&& + 2 \epsilon \sum_{l=1}^N e^{ikl} \sum_{X:m} e(X) 
 \sum_{<j,j+1> \in \boundary ((X \dif M)+l)} 
 | \{j,j+1\} \dif [(X \dif M)+l] \rangle
\eea
Using the change of variables $j \ra j+l$ in the second term, this is 
\bea
&=& 2 \sum_{l=1}^N e^{ikl} \sum_{X:m} e(X) 
 n(X) \, \, |(X \dif M)+l\rangle
\nonumber \\
&& + 2 \epsilon \sum_{l=1}^N e^{ikl} \sum_{X:m} \quad
\sum_{<j,j+1> \in \boundary (X \dif M)} e(X) \,\,
 | (\{j,j+1\} \dif X \dif M)+l\rangle
\eea
In the second sum on $X$ we do a change of variables: 
$X \ra X \dif \{j,j+1\}$. 
Since $<j,j+1> \in \boundary (X \dif M)$ if and only if 
$<j,j+1> \in \boundary (\{j,j+1\} \dif X \dif M)$, we obtain
\bea 
H \Psi_k &=& 
2 \sum_{l=1}^N e^{ikl} \sum_{X:m} e(X) 
 n(X) \, \, |(X \dif M)+l\rangle
\nonumber \\
&& + 2 \epsilon \sum_{l=1}^N e^{ikl} \quad
\sum_{X:m} \quad \sum_{<j,j+1> \in \boundary (X \dif M)} e(X \dif \{j,j+1\})
 \,\, | (X \dif M)+l\rangle
\label{sels}
\eea

The above should be equal to $E(k) \Psi_k$. The eigenvalue now depends on 
$k$. We write $E(k)$ as a Fourier series. Since it equals $2$ 
when $\epsilon=0$, we take the series in the form  
\be
E(k) = 2 + \sum_{s=1}^N e_s e^{iks}
\label{efs}
\ee
So 
\bea
(E(k)-2) \Psi_k &=& \sum_{s=1}^N e_s e^{iks} 
\sum_{l=1}^N e^{ikl} \sum_{X:m} e(X) \, \, |(X \dif M)+l\rangle
\nonumber \\
&=& \sum_{s=1}^N \sum_{l=1}^N 
e_s e^{ikl} \sum_{X:m} e(X) \, \, |(X \dif M)+l-s\rangle
\eea
where we have used the change of variables $l \rightarrow l-s$. 
Define $Y$ by 
\be 
(X \dif M)+l-s=(Y \dif M)+l
\ee
Then solving for $X$ in terms of $Y$ we find 
\be
X = (Y+s) \dif (M+s) \dif M
\ee 
So
\be
(E(k)-2) \Psi_k =
\sum_{s=1}^N \sum_{l=1}^N 
e_s e^{ikl} \sum_{X:m} 
e((Y+s) \dif (M+s) \dif M)
\, \, 
 |(Y \dif M)+l\rangle
\label{sers}
\ee

We now multiply both of \reff{sels} and \reff{sers} by 
$e^{-ikn}$ and sum on $k$. This yields  
\bea
&& 2 \sum_{X:m} e(X) 
 (n(X)-2) \, \, |(X \dif M)+n\rangle
\nonumber \\
&& + 2 \epsilon 
\sum_{X:m} \sum_{j \in \boundary (X \dif M)} e(X \dif \{j,j+1\})
 \,\, | (X \dif M)+n\rangle
\nonumber \\
&=&
\sum_{s=1}^N 
e_s  \sum_{X:m} 
e((Y+s) \dif (M+s) \dif M)
\, \,  |(Y \dif M)+n\rangle
\eea
Thus for sets $X$ such that $X:m$ we have
\bea
&& 2 (n(X)-2) e(X) + 2 \epsilon 
\sum_{j \in \boundary (X \dif M)} e(X \dif \{j,j+1\})
\nonumber \\
&&=
\sum_{s=1}^N e_s \, 
e((X+s) \dif (M+s) \dif M)
\label{se} 
\eea

The smallest $n(X)$ can be is 2. It attains this value for the empty
set and for $X$ of the form $M \dif (M+n)$ for some $n$. 
(These are the sets for which $X \dif M$ is just a translate of $M$.)
We take $e(\emptyset)=1$ and $e(X)=0$ for all other $X$ with 
$n(X)=2$.  It is not clear at first that we can do this, but if we 
can succeed in constructing the eigenfunctions under this condition that
will show that these conditions can be imposed. We note that since we now
have $N$ eigenfunctions, there are $N$ degrees of freedom corresponding 
to their normalizations. The conditions we impose can be thought of 
as fixing these normalizations. 

If $X=M \dif (M-n)$, then 
\be
\sum_{s=1}^N e_s \, 
e((X+s) \dif (M+s) \dif M)
=e_n 
\ee
So for these $X$, eq. \reff{se} becomes
\be
e_n =
2 \epsilon \sum_{j \in \boundary (M-n)} 
e(M \dif (M-n) \dif \{j,j+1\})
\label{fpea}
\ee
For $X$ with $n(X)>2$, we solve eq. \reff{se} for $e(X)$ :
\bea
&&  e(X) = -  { \epsilon \over n(X)-2} 
\sum_{j \in \boundary (X \dif M)} e(X \dif \{j,j+1\})
\nonumber \\
&& + {1 \over 2} { 1 \over n(X)-2} \sum_{s=1}^N e_s \, 
e((X+s) \dif (M+s) \dif M)
\label{fpeb}
\eea

We use $e$ to denote the collection of variables $e(X)$ for $X$ such that 
$X:m$ and $n(X)>2$ and the variables $e_n$. 
These are the unknowns in \reff{fpea} and \reff{fpeb}.
The right sides of \reff{fpea} and \reff{fpeb} define a function $F(e)$, 
and together these two equations can be written as the fixed point equation
$F(e)=e$. 
It is important to ask if the argument of any of the $e(\quad)$ in the 
right sides of these equations can be the empty set or a set of the 
form $M \dif (M+s)$. It is easy to check this does not happen in 
\reff{fpea}, except for the trivial case of $m=1$. And it does not happen 
in the second term in \reff{fpeb} thanks to the constraint $n(X)>2$. 
But it can happen in the first term in \reff{fpeb}.
Since $e(M \dif (M+s))=0$ for $s \ne 0$, these terms drop out of 
\reff{fpeb}. And since $e(\emptyset)=1$, when $X=\{m,m+1\}$ or 
$X=\{1,N\}$ we get a contribution of $-\epsilon/2$. 
In particular, $F(0)$ is not the zero vector in our Banach space.

We first show the fixed point equation has a solution using the norm 
\be
||e|| = \sum_{n=1}^N |e_n| 
+ 2 \sum_{X:m} |e(X)| (n(X)-2) 
\label{norm}
\ee
We will show that with the above norm, 
$F$ is a contraction in a small ball about the origin. 
We have 
\bea 
&&||F(e)-F(e')|| \le 
2 \epsilon  \sum_n \sum_{j \in \boundary (M-n)} 
|e(M \dif (M-n) \dif \{j,j+1\})-e'(M \dif (M-n) \dif \{j,j+1\})|
\nonumber \\
&+&  2 \epsilon \sum_{X: n(X)>2, X:m}  \quad
\sum_{j \in \boundary (X \dif M)} 
|e(X \dif \{j,j+1\}) - e'(X \dif \{j,j+1\})|
\nonumber \\
&+& \sum_{X:n(X)>2, X:m} \sum_s 
|e_s \, e((X+s) \dif (M+s) \dif M)-e'_s \, e'((X+s) \dif (M+s) \dif M)|
\label{threesums}
\eea
The terms in the first sum in the above are the terms one would get 
from the second sum for $X$ with $n(X)=2$. So together these two
sums are 
\bea
&=&  2 \epsilon \sum_{X:m} \quad \sum_{j \in \boundary (X \dif M)} 
|e(X \dif \{j,j+1\}) - e'(X \dif \{j,j+1\})|
\nonumber \\
&+&  2  \epsilon \sum_{X:m} \quad \sum_{j \in \boundary (X \dif M)} 
|e(X) - e'(X)| = 2 \epsilon \sum_{X:m} n(X) |e(X)-e'(X)|
\nonumber \\
\eea
where we have used the change of variables $X \rightarrow X \dif  \{j,j+1\}$.
If $n(X)=2$, then $e(X)=e'(X)$.
For the other $X$, $n(X) \ge 4$ and so $n(X) \le 2 (n(X)-2)$. So the above is 
$\le 4 \epsilon ||e-e'||$.

In the third sum in \reff{threesums},
we drop the constraint $n(X)>2$ and do a change of variables 
$Y=(X+s) \dif (M+s) \dif M$. Then it is 
\bea
\le && \sum_{Y:m} \sum_s 
|e_s e(Y)- e'_s \, e'(Y)|
\nonumber \\
\le&& \sum_{Y:m} \sum_s 
|e_s| \, |e(Y)-e'(Y)|
+ \sum_{Y:m} \sum_s 
|e_s-e'_s| \, |e'(Y)|
\nonumber \\
&\le & \max \{||e||,||e'||\} ||e-e'|| 
\le \delta ||e - e'||
\eea
Thus $F$ is a contraction on a ball of radius $\delta$ about the origin
if $\delta$ and $\epsilon$ are small enough. 
Since $||F(0)||$ is of order $\epsilon$, the above 
estimate also shows that  
$F$ maps the ball $\{e: ||e|| < \delta\}$ back 
into itself if $\delta$ and $\epsilon$ are small enough.

As in the previous section, we can introduce a stronger norm. Let
\be
||e|| = \sum_{n=1}^N |e_n| (K |\epsilon|)^{-w_n} 
+ 2 \sum_{X:m} |e(X)| (n(X)-2)  (K |\epsilon|)^{-w(X)} 
\ee
where $w(X)$ is defined as in the previous section and 
$w_n=w(M \Delta (M+n))$. We emphasize that this is a natural norm in the 
sense that the powers $w_n$ and $w(X)$ are the lowest order in perturbation
theory at which the corresponding terms get nonzero contributions.
We leave it to the reader to check that the preceding estimates go 
through if $K |\epsilon| \le 1$ and $K$ is large enough. It is easy to 
see that $w_1=m$ and the other $w_n$ are even larger. Thus this 
stronger norm shows that the coefficients in the Fourier series of 
the dispersion relation are at least order $\epsilon^{m}$.

We summarize our results on the droplet eigenstates in a theorem. 

\begin{theorem} For the chain with periodic boundary conditions and $N$ 
sites, we consider the sector with $m$ down spins ($0<m<N$).
There is a constant $K>0$ (independent of $N$) 
such that if $K |\epsilon|<1$, then the 
fixed point equation has a solution in the above norm. The $N$ eigenstates
and eigenvalues determined by this solution through \reff{eigenfunction}
and \reff{efs} have the lowest eigenvalues in this sector. 
(Except for $k=0$ in which case it is the second lowest.) 
Letting $E_N(k)$ denote the 
eigenvalue for a chain with $N$ sites corresponding to momentum $k$, 
there are constants $d_s$ for $s \in \zed$ such that 
\be
\lim_{N \rightarrow \infty} E_N(k) = 2 + \sum_{s=-\infty}^{\infty} 
d_s e^{iks}
\label{thmeq}
\ee
The Fourier coefficients $d_s$ are absolutely summable, and 
for $s \ne 0$ the coefficients are of
order at most $\epsilon^m$.
\end{theorem}

The existence of a solution to the fixed point equation implies that 
we have constructed $N$ eigenvalues and eigenfunctions. 
To see that for each $k$ the eigenvalue is the lowest eigenvalue 
in the sector of momentum $k$ (or second lowest if $k=0$) we argue as 
follows. We know the claim is true when $\epsilon=0$. For a finite chain the 
eigenvalues are continuous in $\epsilon$, and so our eigenvalue
can cease being the lowest as $\epsilon$ is increased 
only by crossing another eigenvalue. At the value of $\epsilon$ where such 
a crossing would occur, our eigenvalue would be degenerate. One can then
show that this implies the solution of the fixed point equation is not
locally unique, contradicting the contraction mapping theorem. 
(More details may be found in \cite{DKa}.) 

For a finite chain the eigenvalues $E_N(k)$ are only defined by a 
finite set of values which depends on $N$. But we can use use \reff{efs}
to extend the definition to all $k$ and so make sense of the limit 
in \reff{thmeq}.
The existence of the limit as $N \rightarrow \infty$ of the Fourier 
coefficients $e_s$ follows by standard arguments. (See \cite{DKa} for 
similar arguments.) 

\section*{Acknowledgements}
The author would like to thank the Institute des Hautes \'Etudes 
Scientifiques where this work was done for their support and 
hospitality.
This work was supported by the National Science Foundation (DMS-0201566).

\bigskip

\noindent

\bibliographystyle{amsalpha}

\end{document}